\documentclass[11pt,a4paper]{article}

\usepackage[utf8]{inputenc}
\usepackage[T1]{fontenc}

\usepackage[margin=1in]{geometry}
\usepackage{setspace}
\usepackage{microtype}

\usepackage{booktabs}
\usepackage{array}
\usepackage{tabularx}
\usepackage{multirow}

\usepackage{enumitem}

\usepackage{xcolor}
\usepackage[most]{tcolorbox}

\newtcolorbox{calloutbox}[1][]{
  colback=gray!8,
  colframe=gray!40,
  fonttitle=\bfseries,
  title=#1,
  boxrule=0.5pt,
  arc=3pt,
  left=8pt,
  right=8pt,
  top=6pt,
  bottom=6pt,
  breakable
}

\usepackage[hyphens]{url}

\usepackage[colorlinks=true,linkcolor=blue,citecolor=blue,urlcolor=blue,breaklinks=true]{hyperref}

\title{Beyond the Binary: A nuanced path for open-weight advanced AI}

\author{
  Beng\"{u}su \"{O}zcan,\quad Alex Petropoulos,\quad Max Reddel \\[4pt]
  \textit{Centre for Future Generations (CFG)} \\
  \textit{Brussels, Belgium}
}

\date{July 30, 2025}

\begin{document}
\maketitle

\begin{abstract}
Open-weight advanced AI models---systems whose parameters are freely available for download and adaptation---are reshaping the global AI landscape. As these models rapidly close the performance gap with closed alternatives, they enable breakthrough research and broaden access to powerful tools. However, once released, they cannot be recalled, and their built-in safeguards can be bypassed through fine-tuning or jailbreaking, posing risks that current governance frameworks are not equipped to address.

This report moves beyond the binary framing of ``open'' versus ``closed'' AI. We assess the current landscape of open-weight advanced AI, examining technical capabilities, risk profiles, and regulatory responses across the European Union, United States, China, the United Kingdom, and international forums. We find significant disparities in safety practices across developers and jurisdictions, with no commonly adopted standards for determining when or how advanced models should be released openly.

We propose a tiered, safety-anchored approach to model release, where openness is determined by rigorous risk assessment and demonstrated safety rather than ideology or commercial pressure. We outline actionable recommendations for developers, evaluators, standard-setters, and policymakers to enable responsible openness while investing in technical safeguards and societal preparedness.
\end{abstract}


\section{Executive Summary}
\label{sec:executive-summary}

Open-weight advanced AI models---systems whose parameters are freely available for download and adaptation---are reshaping the global AI landscape. Their proliferation is accelerating both opportunity and risk, as well as stretching existing governance frameworks to their limits.

These models are rapidly closing the performance gap with closed alternatives. They enable breakthrough research, broaden access to powerful tools, and drive innovation globally. But once released, they cannot be contained. Others can fine-tune or repurpose them to bypass built-in safeguards---posing systemic risks that neither existing regulations nor safeguards can contain.

This report rejects the binary of ``open'' versus ``closed.'' Instead, it proposes a tiered, safety-anchored approach to model release. Openness should be based on rigorous risk assessment and demonstrated safety---not ideology, precedent, or commercial pressure. Models should be opened only when there is credible evidence that risks can be effectively mitigated.

But we have to think beyond technical safeguards. Building public understanding, institutional literacy, and responsive governance capacity is essential to ensure that openness---when justified---translates into safe and accountable deployment.

To advance this agenda, the report outlines practical recommendations for key actors which we broadly group into three categories (see Table~\ref{tab:actors}).

\begin{table}[htbp]
  \centering
  \caption{Key actor categories and their roles in open-weight advanced AI governance.}
  \label{tab:actors}
  \begin{tabularx}{\textwidth}{>{\bfseries\raggedright\arraybackslash}p{2.2cm} X X X}
    \toprule
    & \textbf{Builders \& Enablers} & \textbf{Evaluators \& Standard-Setters} & \textbf{Implementers \& Enforcers} \\
    \midrule
    Who &
    AI developers, funders, investors, open-source communities &
    AI Safety Institutes, experts and specialists in AI or adjacent domains, international oversight bodies &
    Governments, security agencies, public institutions \\
    \midrule
    Role &
    Develop advanced AI systems or provide the capital, infrastructure, and platforms that enable their creation, distribution, and adoption---open or closed. &
    Assess systems, define risks and safety thresholds, and develop frameworks both in AI and adjacent fields like biosecurity, cybersecurity, and more. &
    Deploy safeguards, regulate model use, invest in resilience, and translate safety standards into policy and societal preparedness. \\
    \bottomrule
  \end{tabularx}
\end{table}

These interventions are not about restricting open innovation---they're about securing it. The goal is to enable openness in advanced AI where risks can be responsibly managed, and to pair that with concrete investments in public and institutional readiness. Without the ability to monitor, evaluate, and respond to misuse, even well-intentioned openness can undermine safety.

\begin{calloutbox}[What are model weights?]
Model weights~\cite{ref1} are the fundamental numerical parameters that define a model's behaviour by adjusting how it prioritises different aspects of its input data. Because weights capture what the model has learned, open-weight models---where these weights are publicly released---are central to debates on openness in AI, as they allow others to reuse, modify, or adapt the models.
\end{calloutbox}

\section{Introduction}
\label{sec:introduction}

As more capable AI models are released seemingly every other week, there's a narrative forming of a cat and mouse game between open and closed approaches. This plays out on two levels: the technical aspect where the open-weight models rapidly catch up with the capabilities of closed systems in only a few months, and the broader, decades-old tension~\cite{ref2} between open-source and proprietary development models, now emerging in the AI landscape with much higher stakes.

While capabilities race ahead, governance is still trailing behind---particularly when it comes to how, when, and whether to release powerful models with open weights. Openness is often treated as a goal in itself, even when the conditions for doing so safely are unclear or unproven.

This report proposes a different path forward: a tiered approach to open-weight release, where decisions about model openness are based not on ideology or precedent, but on demonstrated safety and contextual risk. Such a framework would allow more capable models to be shared under the right conditions---enabling scientific and public value---while withholding or restricting access when credible threats to misuse exist.

To support this shift, we begin by assessing the current landscape of open-weight advanced AI, including the technical, regulatory, and institutional gaps that define it. We then outline actionable recommendations across key stakeholder groups that together could make selective openness both safer and more sustainable. If implemented, these measures can help ensure that today's choices do not foreclose the possibility of safe openness tomorrow---when models may be both significantly more capable and harder to govern.

\section{Current Landscape of Open-Weight Advanced AI}
\label{sec:landscape}

\subsection{A Rapidly Changing Frontier}
\label{sec:frontier}

The pace of advanced AI model releases---highly capable general-purpose AI models~\cite{ref3}---continues to accelerate dramatically. In early 2025, OpenAI~\cite{ref4} and DeepSeek~\cite{ref5} each announced comparable models highly capable of software. OpenAI's model was closed-source, while DeepSeek's was open-weight. Just months later, OpenAI~\cite{ref6} and Anthropic~\cite{ref7} released new closed models that significantly outperformed their predecessors. Shortly after that, DeepSeek rolled out a major update~\cite{ref8} to its R1 open-weight model, bringing its performance back in close range with the state-of-the-art closed models. These developments suggest a fast-paced catch-up dynamic, with open-weight and closed models advancing in a tight loop.

\begin{calloutbox}[1. What exactly do we mean by ``open-weight'' models?]
Advanced AI models consist of several key components: training data, code architecture, and model weights. Often, models are claimed to be released ``openly'' when only their weights are shared, while other critical components remain withheld. The weights define the model's behaviour as the result of a very resource-intensive training process, as a fundamental component. With access to weights, users can fine-tune models extensively for various purposes, making this component particularly significant for governance considerations. In this piece, we use ``open-weight'' to refer specifically to models whose parameters are publicly downloadable and can be freely modified and fine-tuned, highlighting the unique governance challenges they present.
\end{calloutbox}

\begin{calloutbox}[2. Why don't we simply call them ``open-source AI''?]
In scientific fields, open source traditionally means full transparency, where code, data, and methodologies are publicly available for collective improvement. In software specifically, open source also enables redistribution, modification, and collaborative development through standardised licensing frameworks~\cite{ref9} (like MIT, Apache, or GPL) that primarily address copyright and redistribution rights rather than restricting specific applications. The Open Source Initiative recently released~\cite{ref10} a definition for open-source AI that maintains these comprehensive standards across both scientific and software dimensions. However, in advanced AI the term ``open'' has become considerably fuzzier, because many ``open'' releases exclude key components such as training data or full code. Therefore, ``open'' or ``open-source'' advanced AI often overstates the actual degree of transparency.
\end{calloutbox}

\begin{calloutbox}[3. Why does this distinction matter?]
From a transparency perspective, users cannot reproduce models from scratch without access to training code and data. While AI developers may have legitimate reasons for not sharing all model components such as protecting competitive advantages or privacy concerns, they should be more precise on what model components they are sharing in an ``open'' release. From a governance and risk management perspective, access to weights is key to repurpose a model, which makes this component particularly relevant. While licensing terms of open-weight models might attempt to prevent misuse of these models, their enforceability is often limited---as recently demonstrated~\cite{ref11} when Meta's open-weight models were allegedly used for Chinese military purposes despite explicit prohibitions in their licence terms. This enforcement gap makes the technical accessibility of weights, rather than legal restrictions, a more pressing governance concern. Efforts~\cite{ref12} are underway to define model openness based on access to weights, code, and data, and licensing restrictions. While some argue that this compromises~\cite{ref13} the very definition of openness, a nuanced approach in definitions seems necessary in order to assess the current landscape.
\end{calloutbox}

\subsection{Benefits in the Balance}
\label{sec:benefits}

The capability gap between open and closed models, while still present, has recently narrowed. A 2024 report by Epoch AI found~\cite{ref14} that open-weight models lag behind the most advanced closed models by approximately 5 to 22 months depending on the task, and they expect this gap to close rapidly. Since Epoch's publication, new model releases have sent mixed signals---with DeepSeek indeed closing this gap, and Meta's Llama~4 underperforming expectations---but continue to gain significant popularity. Hugging Face~\cite{ref15}, a central repository where open models are typically published, hosts more than 200,000 text-generation models. As of July 2025, 5 of the top 10 most downloaded text generation models are general-purpose advanced AI models released within one year, demonstrating how quickly these models have been diffusing.

This diffusion suggests potential benefits, though specific studies on the economic impact of open-weight advanced AI models are limited. We can infer possible value from related research: broader open science initiatives generated an estimated \texteuro65--95 billion annually for EU GDP according to a 2021 European Commission study~\cite{ref16}, while a 2024 econometric study~\cite{ref17} found that open-source software is expected to contribute approximately 2.2\% to national GDPs globally in the long run. In particular, an annual report~\cite{ref18} by GitHub, a widely used platform by developers, documented nearly 150,000 open generative AI projects on their platform in 2024, representing 98\% growth compared to 2023. The report also highlights a surge in contributors from Asia, Africa, Latin America and many emerging economies compared to 2023, attributing this indirectly to availability of open AI tools.

While comprehensive evidence on open-weight advanced AI's specific contributions to the economy and society remains limited, these parallel trends suggest similar positive impacts may emerge as these models are adopted more.

\subsection{Risks in Plain Sight}
\label{sec:risks}

However, as these powerful capabilities diffuse widely, so do their potential risks. Our previous research has shown that advanced AI models exhibit dual-use capabilities~\cite{ref19} that can pose increasingly serious risks as these models grow more capable. Open-weight releases can lower~\cite{ref20} the barrier to misuse for well-resourced actors. Moreover, once these models are released, they cannot be recalled or contained.

\begin{calloutbox}[Why is it so hard to release open-weight AI models safely?]
Open-weight AI models come with no undo button. Once publicised, their weights---essential components that determine the model's behaviour---can be downloaded, copied, and redistributed freely and indefinitely. That raises critical questions: if these weights are just arrays of numbers, what makes them potentially dangerous once they are released? Can we not build some sort of safety mechanism around these numbers to prevent misuse?

The answer lies in the model's architecture and how easily its behaviour can be changed. When model weights are openly available, individuals can bypass built-in safeguards---a process known as jailbreaking---or fine-tune the model on new data to alter its outputs with minimal cost and effort. Adversarial evaluations~\cite{ref21} of recent models like DeepSeek's R1 demonstrate that novel techniques to circumvent safety guardrails are evolving rapidly, keeping pace~\cite{ref22} with the safeguards of the most recent advanced AI models.

Importantly, risk isn't limited to intentional misuse. A recent study~\cite{ref23} on safety training durability shows that even well-meaning users can accidentally weaken safety systems during fine-tuning. Small, seemingly harmless changes can result in models that behave in undesired ways---yet these failures may go unnoticed, with standard evaluations often failing to flag such regressions. This makes unintended misuse a serious concern, especially as proliferation of misuse can happen almost immediately~\cite{ref24} after release, as demonstrated in several early examples.

While we have not yet witnessed widespread catastrophic misuse of open-weight models, this absence should not be mistaken for evidence of safety. As the capabilities of these models advance rapidly, waiting for clear evidence of harm before acting could be a costly mistake. Policy and safeguards must evolve just as swiftly, building societal resilience and requiring safety assurances for powerful open releases before the window for meaningful intervention closes.
\end{calloutbox}

The International AI Safety Report~\cite{ref25} (released in January 2025, coordinated by the UK Government with contributions from expert delegations across 30 countries) dedicates a chapter to a key risk of open-weight advanced AI models. There is broad consensus that future models are highly likely to significantly assist motivated users with average domain-specific expertise across various threat domains, ranging from novel pathogen creation to automated cyberattacks. Critically, the current lack of preemptive tracking or evaluation means we may only realise we've crossed a dangerous threshold after such models have already been released and misused.

Given that these risks are primarily tied to vast and emerging model capabilities, let's pause and delineate ``open-weight advanced AI'' in consideration from the broader umbrella of ``open-weight AI models.'' To illustrate the lack of clarity in the landscape, let's look more closely at the models we mentioned earlier.

Among the top 10 most downloaded open-weight text generator models on Hugging Face, some are ${\sim}400$ times larger in size than others, represented with parameter count. While model size is no longer a perfect measure~\cite{ref26} of an AI system's capabilities or risks, it still presents~\cite{ref27} a proxy: substantially larger advanced AI models are often more capable than smaller ones. This disparity amongst the top 10 most downloaded open-weight models indicates substantially different capability and risk profiles that shouldn't be evaluated under a single ``open-weight'' umbrella.

The landscape is further complicated by the divergent aims of the developers. Looking at the same subset of 10 models, we see evidence of this: one small language model~\cite{ref28} seems to merely aim at optimising its cost and performance, another~\cite{ref29} is released by a company explicitly working towards artificial general intelligence (AGI), while another model released by a company~\cite{ref30} with the mission of opening the most cutting-edge AI models for all. In this fragmented ecosystem, it is increasingly important for policymakers and governance professionals to delineate open-weight advanced AI models from other AI systems, recognising their unique benefits and challenges rather than applying one-size-fits-all approaches---and to pursue an openness approach that is grounded in demonstrated safety.

\section{Governance Gaps and Global Responses}
\label{sec:governance}

Currently, there is no commonly adopted industry standard or universal framework for assessing under what conditions and classifications advanced AI models should be released as open-weight to maximise benefits while minimising risks. Regulatory frameworks that address both advanced AI capabilities and open-weight distribution remain a niche, leaving significant governance gaps.

\subsection{European Union}
\label{sec:eu}

The EU AI Act stands out as one of the most comprehensive regulatory approaches addressing both advanced AI capabilities and open-weight models. In July 2025, the final version of the General-Purpose AI Code of Practice~\cite{ref31} was released as a voluntary guideline to help developers comply with the Act's requirements. All general-purpose models, whether open- or closed-weight, are advised to adopt core technical safety measures before release. While the global response---especially from open-weight developers---remains uncertain, early momentum is visible among EU players. Mistral, a leading French AI company known for its open-weight models, is reported~\cite{ref32} to have committed to the Code of Practice. The Code also acknowledges inherent challenges, such as monitoring misuse after release. Given the difficulty of monitoring open-weight models post-deployment, it instead encourages developers to focus on pre-release safeguards, clear documentation, and communicating risks to downstream users.

\subsection{The United States}
\label{sec:us}

The United States presents a shifting regulatory landscape. The former US President Biden took two notable actions targeting advanced AI models: Executive Order 14110~\cite{ref33} (2023) established safety standards and reporting requirements for advanced AI developers, but this was rescinded by President Trump in January 2025. Currently, NIST's AI Risk Management Framework~\cite{ref34} provides voluntary guidelines for managing AI risks that could apply to open-weight models, but no enforceable regulatory framework specifically governs their development or deployment. While the United States' regulatory stance on open-weight advanced AI currently remains uncertain, and might be a part of their forthcoming AI Action Plan~\cite{ref35}, a recent public statement~\cite{ref36} by the Senior White House Policy Advisor on Artificial Intelligence suggests growing support for the U.S.\ to position itself as a global leader in open-weight AI development.

\subsection{China}
\label{sec:china}

In China, where DeepSeek's emergence caught~\cite{ref37} global attention, the Interim Measures for the Management of Generative AI Services~\cite{ref38} (2023) applies to both open and closed models; however these regulations focus primarily on conventional software governance issues like content control, IP protection, and user rights with less emphasis on the unique challenges posed by advanced AI systems such as conducting tests for emergent behaviours.

\subsection{The United Kingdom}
\label{sec:uk}

The United Kingdom has taken a proactive stance on advanced AI safety by hosting the inaugural AI Safety Summit~\cite{ref39} and establishing the AI Safety Institute~\cite{ref40}---renamed to ``AI Security Institute'' as of February 2025---to evaluate advanced models for unique risks. While the UK lacks a targeted framework for open-weight advanced AI, an upcoming AI bill~\cite{ref41} expected this year will likely mirror aspects of the EU AI Act in addressing these emerging challenges.

\subsection{International Landscape}
\label{sec:international}

These national and EU regulations show varied approaches to advanced AI governance. Similarly, at the international level, consensus on open-weight model governance does not exist, just as there are no established common standards for advanced AI safety. The most recent major international forum for nations to forge a shared path for advanced AI safety and governance was the AI Action Summit~\cite{ref42} which took place in Paris in February 2025, alongside which CFG published policy recommendations~\cite{ref43} to establish a consensus on the risks of advanced AI and initiate an effort to develop risk standards. However, the summit did not yield a unified, multilateral approach for advanced AI safety, perceived~\cite{ref44} as a missed opportunity by some experts. The summit resulted with a declaration~\cite{ref45} with 64 signatories, missing the US and the UK, which endorses the development of open source AI for the public, without addressing the nuances of advanced AI and other broader AI applications.

In the absence of a standard approach, examining the safety practices of individual advanced AI developers can reveal how they approach these concerns internally. CFG's 2024 assessment~\cite{ref46} found significant disparities across companies---some had implemented robust protocols, while others lacked structured safety frameworks altogether. More recent assessments continue to support this uneven picture. Over time, some AI developers have updated their frameworks, like Meta~\cite{ref47}, while others have still not published extensive frameworks comparable to their peers, like Mistral providing only an input moderation framework~\cite{ref48} without further capability testing. Even though these efforts are fragmented, collaboration~\cite{ref49} between dedicated AI institutes, such as the UK AISI or the EU AI Office, and these developers might lay the building blocks for industry standards.

However, these efforts remain fundamentally voluntary and company-dependent. For instance, Meta recently released~\cite{ref50} their most advanced series of AI models with open-weights, yet provided no public assessment against their own safety framework, which itself includes no commitment to public disclosures. As capabilities advance, this governance gap may become increasingly concerning.

Today, an AI model considered dangerous for open-weight release by one company can be freely published by another with minimal evaluation. The lack of transparency coupled with the absence of common standards means that, even when safety frameworks exist, there's no way to verify either their implementation or effectiveness. Without addressing this fundamental accountability challenge, even well-intentioned safety efforts by individual developers may prove insufficient against the global landscape of AI development.

\section{The Very Near Future of Open-Weight Advanced AI}
\label{sec:near-future}

The trajectory of open-weight advanced AI development suggests that increasingly capable models will continue to be released~\cite{ref51} in the near future. At the same time, risk assessments by individual developers reveal substantial inconsistencies that are likely to persist as this acceleration continues---leaving the governance landscape fragmented even while capabilities converge.

The underlying drivers of openness vary. While ideological motivations like embracing open science and democratising access to technology might play a role, specific strategic considerations might also motivate developers to release open-weight advanced AI models. These include commoditising~\cite{ref52} certain AI capabilities to encourage third-party adoption or indirectly boost demand for other products, and the belief that public scrutiny surfaces~\cite{ref53} safety issues more comprehensively than internal testing alone.

Trade-offs between developer standards and evolving national strategies are likely to shape the trajectory of advanced open-weight AI models. These conversations are evolving in real time, but are already widespread:

\begin{itemize}[leftmargin=*]
  \item In China, we see major organisations like Alibaba and DeepSeek primarily pursuing~\cite{ref54} open-weight rather than closed strategies, with plans~\cite{ref55} to release more capable models in the near future. However, questions~\cite{ref56} remain about whether the Chinese government might disapprove of this trajectory, constraining companies' open release policies to protect strategic national interests.

  \item Mistral AI, the leading advanced AI developer in France, recently reconfirmed~\cite{ref57} its strong commitment to open-weight models, positioning this approach as enabling France's aim~\cite{ref58} to gain sovereignty in AI through international collaborations.

  \item The US is currently drafting its national strategic AI plans, and many observers expect it to adopt a pro-open approach. However, the US might also try to maintain~\cite{ref59} its edge on advanced AI. This suggests a potential ``tier-based'' openness policy, where models up to certain capability thresholds may be encouraged to be open-weight, while the most advanced models that represent strategic technological advantages might remain protected as national security assets. This approach is reflected by various AI developers.

  \item Meta has consistently maintained an ideological stance favouring open-weight releases amongst the major advanced AI developers in the US, releasing~\cite{ref60} their most advanced model Llama~4 with open weights a few months ago.

  \item OpenAI, which has not opened any of its models since GPT-2~\cite{ref61} in 2019, has announced~\cite{ref62} plans to release an open-weight model in the near future.
\end{itemize}

This rapid proliferation collides with uneven safety standards and practices across the industry. Anthropic, a closed-model developer, applies an internal capability scale and has flagged~\cite{ref63} its newest model, Opus~4, carries potentially high CBRN risk. The specific tests and thresholds remain undisclosed, but the judgment appears driven by the model's ability to assist in generating or enabling CBRN threats. Meanwhile, DeepSeek's open-weight R1, updated in May 2025, outperforms Opus~4 on aggregate capability benchmarks reported~\cite{ref64} in July 2025. Although those benchmarks do not align exactly with Anthropic's undisclosed risk metrics, the contrast shows how one company can withhold a model for safety reasons while another releases a functionally similar---indeed more capable---system with minimal constraints, producing a patchwork of risk exposure.

OpenAI's recent actions provide a case study in this tension between openness and safety. When the company first announced plans for a forthcoming open-weight model, they added that they would test this model against their internal Preparedness Framework~\cite{ref65}---a protocol designed to assess risks unique to general-purpose advanced AI, such as assisting with cyberattacks or bioweapon development. Recently, OpenAI announced~\cite{ref66} that they would delay this model's release, citing safety concerns. OpenAI's decision may set an important precedent---not only for evaluating advanced models prior to release, but also for applying structured safety testing specifically to an open-weight model, signalling a growing norm of pre-deployment risk assessment. However, this remains a voluntary effort based on an internal, non-standardised framework with no external oversight. Taken together with the Anthropic--DeepSeek contrast, OpenAI's decision highlights how well-intentioned but ad-hoc assessments cannot ensure system-wide assurance when capabilities are converging.

This overall trajectory reveals several critical questions we must address: how to maximise benefits while effectively managing risks, how to prepare for unforeseen consequences despite precautions, and what approaches might yield more effective technical safeguards. While there are no definitive answers to these complex challenges, the path forward requires balanced, multi-faceted strategies. Open-weight innovation is only as inclusive as its safeguards---without deliberate investment in safety, accessibility risks turning into exposure. In the following section, we explore concrete approaches to meet these challenges head-on, outlining both immediate steps and long-term frameworks to help ensure that open-weight AI delivers on its promise while minimising potential harms.

\section{A Nuanced Approach: How to Have Safe Advanced AI, and Open It Too}
\label{sec:approach}

After examining the current landscape and expected trajectories of open-weight advanced AI, the evidence reveals both substantial benefits that should be preserved and risk factors that may evolve as new capabilities emerge. We need to move beyond the binary of `open versus closed'---what's needed now is a coordinated, multi-layered framework where every stakeholder has a role in making openness not simply open, but safe. A tiered approach, where openness is granted in proportion to demonstrated safety, can offer that structure.

We recognise that this isn't straightforward. It demands targeted interventions from both policymakers and technical researchers, addressing gaps in governance mechanisms and the need for technical safeguards. With this in mind, we offer concrete recommendations across complementary domains. While not being exhaustive, policy interventions that enable these and promote technical work that is necessary to achieve these can enable more advanced open-weight models to be released gradually and responsibly. Importantly, these measures could strengthen the overall safety ecosystem for advanced AI development more broadly, addressing the critical gaps in advanced AI safety~\cite{ref67} that we already observe today.

Below, we introduce these measures along with the specific actions that relevant stakeholders should take to implement them effectively.

\subsection{Establish and Adopt Standards for a Tiered Open-Weight Release}
\label{sec:standards}

Developing clear, rigorous and enforceable standards is what makes a tiered approach to open-weight model release feasible---allowing openness only when it can be demonstrated that release is sufficiently safe. Ideally, such standards would be codified through formal global mechanisms---like ISO certifications---but these processes are sluggish in comparison to fast-evolving advanced AI capabilities. In the meantime, we suggest taking meaningful intermediate steps.

The most immediate priority is to understand what open-weight models can actually do when evaluated against specific threat scenarios. This knowledge gap must be addressed urgently, as model capabilities continue to outpace our current risk assessment frameworks.

\subsubsection{AI Safety Institutes}
\label{sec:aisis}

AI Safety Institutes (AISIs) are uniquely positioned to lead efforts on developing risk thresholds and clearly documented testing methodologies, creating a foundation for making informed decisions on open-weight releases. The UK AI Security Institute's safety case~\cite{ref68} approach provides an example of a structured framework that combines precise safety claims, supporting evidence, and logical arguments to demonstrate that an AI system meets defined safety standards---particularly valuable when assessing whether open-weight releases meet appropriate safety thresholds. The EU AI Office's newly forming network of model evaluators~\cite{ref69}, though announced as to primarily focus on regulatory enforcement, could extend beyond its primary regulatory enforcement focus to develop specific frameworks for open-weight models, especially to ensure that the EU AI Act does not create a two-tier system where closed models face rigorous scrutiny while similarly capable open-weight models are released with minimal oversight.

It is unclear whether all AISIs will prioritise these efforts similarly as each institute operates~\cite{ref70} with distinctly different priorities and mandates. The UK has demonstrated strong dedication to frontier AI safety, but Singapore's AISI appears more focused on application-specific outcomes, and the future direction of the American AISI under the new administration remains uncertain~\cite{ref71}. Given these divergent national priorities, international collaboration between AISIs becomes even more crucial. After all, advanced open-weight models can be downloaded and deployed anywhere, including in countries without frontier AI development capabilities, so the problem of AI risk is inherently global. This demands some level of aligned risk assessment across jurisdictions. The International AI Safety Network~\cite{ref72} holds significant promise in this regard. By coordinating expertise and resources across institutes, it could help establish common standards and evaluation practices, benefiting both open and closed models. International collaboration will be essential to create robust, globally applicable safety frameworks and to ensure that the oversight of advanced AI models is not fragmented, but resilient and adaptive across borders.

\subsubsection{Experts and Specialists in AI-Adjacent Domains}
\label{sec:domain-experts}

Domain experts, especially in areas where misuse potential is acute such as chemical weapon proliferation or critical infrastructure management, should be encouraged to take an active role in building institutional knowledge on how AI misuse risks are understood, evaluated, and governed. Their engagement should go beyond one-off consultations, and instead work in tandem with AISIs in order to develop robust threat models, benchmarks and evaluation standards. Given the breadth and complexity of AI capabilities across scientific domains, a more granular delineation of risks and application contexts is essential to bridge technical realities with clear guidance for policy and governance decisions at the right level of abstraction.

An illustration of this need emerged in two recent reports from the National Academies of Sciences~\cite{ref73} and the RAND Corporation~\cite{ref74}, which examined biological misuse risks from different vantage points. The National Academies focused on biological design tools---specialised AI-enabled systems such as protein engineering platforms---not currently possessing the capacity to autonomously produce pandemic-level pathogens. General-purpose AI models and scientific LLMs are not examined in depth; instead, the report briefly acknowledges that such models may pose potential concerns in the future. In contrast, the RAND report mainly evaluated the capabilities of the most recent advanced LLMs for misuse, warning that they are rapidly approaching expert-level performance in biology. RAND examines in detail the technical reasons and developments that increase the plausibility of real-world misuse by capable actors rapidly. Both reports agree that today's AI systems do not yet enable fully automated biothreat creation and still require substantial bioscience expertise---on the face of it, this is a reassuring consensus. However, the two reports prioritise different parts of the AI-biology ecosystem and present the risks with very different levels of urgency. This contrast underscores how wide the threat landscape can be---even within a narrow domain---and highlights the importance of mapping individual risk areas clearly so that policy responses stay proportionate. In practice, these contrasting framings can imply different sets of actions---where one benefits from continued monitoring more than another which might need more proactive prevention methods---shaping how institutions define and address AI-enabled biological threats.

Because different areas of application pose distinct types of risks and decision-making challenges, governance structures must be equipped to respond with appropriate nuance and contextual understanding. One promising approach would be to embed scientific governance expertise into the emerging AI safety infrastructure. International bodies such as the Biological Weapons Convention Implementation Support Unit and the Organisation for the Prohibition of Chemical Weapons (OPCW) offer models for systematically integrating technical expertise into governance---such as the OPCW's Scientific Advisory Board~\cite{ref75}, which regularly assesses advances in chemistry that could impact treaty compliance, including new synthesis technologies and dual-use chemical manufacturing trends. Building similar structures within AISIs could help ensure that AI safety evaluations remain grounded in evolving technical realities rather than abstract assessments.

Alternatively---or in parallel---governments could build secure national frameworks that pair AISIs with national security agencies. Through this model, AISIs would collaborate closely with relevant intelligence and biosecurity bodies, using classified data and operational expertise to shape evaluations. AISIs would then act as trusted intermediaries, translating sensitive findings into structured guidance for international coordination and regulatory development, without exposing critical vulnerabilities publicly.

In either approach, the goal is clear: domain expertise must move from the margins to the centre of AI safety governance, helping to institutionalise a mature and trusted system for evaluating emerging threats.

\subsection{Open-Weight AI Developers and Providers}
\label{sec:developers}

Open-weight developers must embed accountability into their release processes, going beyond licensing terms as those often face significant enforcement challenges. When Meta's open-weight models were allegedly used~\cite{ref76} for military applications despite explicit prohibitions in its licensing terms, it became clear that legal protections offer little practical security once model weights are publicly available. Adequate accountability should involve adherence to evolving, state-of-the-art risk-assessment standards before releasing any model.

This does not imply that existing models must be withdrawn---many may still be considered low-risk under a credible standard. However, as models grow more capable, developers should be ready with a staged-release playbook. For most systems, open access may remain appropriate; but for frontier models that exceed defined safety thresholds yet offer clear scientific benefit, restricted channels such as verified-access programs can be used to grant downloads only to credentialed researchers. These access frameworks, paired with adequate risk categorisation, can provide stronger safeguards---drawing from practices in biomedical data repositories like dbGaP~\cite{ref77} or the UK Biobank~\cite{ref78}, which ensure that only trusted users gain access and are held to appropriate security standards to protect against leaks or misuse.

To complement more restrictive frameworks while supporting academic and research use, sandboxed environments can offer a practical middle ground. Such environments allow researchers and developers to interact with models without directly accessing the raw weights. Sandboxes~\cite{ref79} are commonly used to isolate unstable instances of a model from the main system---whether for further testing or to safeguard the privacy of user inputs, such as proprietary datasets. Academic institutions~\cite{ref80} or developer platforms like GitHub~\cite{ref81} already support such approaches, enabling meaningful experimentation and research while maintaining important security boundaries. Together, controlled distribution and sandboxed access offer differentiated exposure: giving high-risk capabilities only to vetted users while allowing broader, lower-risk experimentation through limited-access channels. Although these mechanisms stop short of offering fully open access, they can significantly reduce exposure risks while still fostering productive engagement with advanced model capabilities.

It's also important to recognise that building and maintaining these access structures involves real operational demands---including ongoing investment, monitoring, and infrastructure support. Developers who remain committed to the ideals of openness should be prepared to invest in these systems. In parallel, they can also support the broader ecosystem by promoting the adoption of existing open-weight models, helping communities integrate and use them responsibly. This focus on supporting adoption---rather than releasing increasingly powerful models simply in pursuit of an open-access ideal---can reinforce both safety and innovation.

At the same time, it is worth asking: will capability gains eventually push some models beyond the scope of any safe openness tier---leaving most of the world without access to them? This is not just a hypothetical scenario, but a natural consequence of a robust risk mitigation policy that may come to represent the safest course of action for our societies---especially if supported by multilateral consensus among governments, safety institutions, and technical experts.

Of course, tiered access frameworks can also evolve. For instance, more advanced tracing, watermarking, or provenance technologies could eventually enable the release of more capable models in ways that are safer and more accountable. While current techniques may still be circumvented, this is precisely why we need more research and investment into strengthening these safeguards---an area the next section explores in more detail.

\subsection{Invest in Technical Safeguards for Open-Weight Models}
\label{sec:safeguards}

To accommodate increasingly advanced open-weight models while managing their risks, developers must go beyond broad commitments to openness and invest in safeguards that align with emerging risk standards. Yet building guardrails that preserve accessibility while preventing misuse is far from straightforward. As capabilities grow, traditional safety measures often fall short, and misuse risks increase in both scale and subtlety. Meeting this challenge requires targeted research into interventions designed specifically for open-weight release---ones that remain effective even when models are widely distributed and potentially modified.

\subsubsection{Open-Weight AI Developers}
\label{sec:dev-safeguards}

As the first line of defence in open-weight deployment, developers must look beyond access controls to the deeper challenge of what models can do once accessed. Embedding safeguards directly into model architecture is essential to reduce risks of malicious repurposing or escalation.

Recent approaches like Tampering Attack Resistance (TAR~\cite{ref82}) show promise by adding defensive layers that resist removal or disabling while maintaining effectiveness for intended uses. Similarly, emerging methods like Deep-Lock~\cite{ref83} might enable parameter-level encryption, ensuring that models can be modified only when the appropriate decryption key is provided by authorised users. Additionally, machine unlearning~\cite{ref84} represents an important frontier: a method that could selectively erase high-risk behaviours or knowledge from a trained model, while preserving useful capabilities.

These approaches offer promising directions for securing open-weight releases at the model level---but they remain early-stage and technically demanding.

Real-world viability will require~\cite{ref85} sustained research, performance evaluations, and threat modelling, particularly given that even heavily protected closed models remain susceptible to jailbreaks. Without greater incentives, clearer deployment norms, or structured support for safety-focused innovation, such protections may remain underutilised. For developers committed to releasing open-weight models, these responsibilities must be embraced as part of the openness itself. Making powerful models broadly accessible is a meaningful contribution to research and innovation---but doing so safely requires care, infrastructure, and discipline equal to that ambition.

Even so, it exemplifies the kind of proactive, safety-oriented research we need more of: not just identifying misuse risks, but designing interventions to neutralise them. Unlearning is just one such avenue.

\subsubsection{Research Funders, Investors, and Open Source AI Communities}
\label{sec:funders}

While developers hold primary responsibility for building safeguards into open-weight models, tackling misuse risks is too complex to be solved by any single organisation or research direction. The techniques highlighted above---TAR, Deep-Lock, and machine unlearning---point to promising technical defences, but they remain early-stage and carry considerable uncertainties. For example, questions remain~\cite{ref86} about whether unlearning can be made scalable, resistant to reactivation, and robust against adversarial retraining. Similarly, recent findings~\cite{ref87} show that even well-meaning users can inadvertently bypass safety guardrails---often without current evaluations detecting it. These limitations reveal how unreliable today's safeguards can be and underscore the urgent need for new strategies that are not only testable and robust, but also resilient to evolving misuse tactics.

Given these uncertainties, advancing the next generation of safeguards will require broader engagement across the ecosystem. Public and philanthropic research funders, academic institutions, and open research collectives are pivotal in this effort---supporting the foundational work needed to design technical interventions that remain effective under open-weight conditions. Beyond access governance, there is growing demand for research that directly explores how to limit high-risk capabilities post-release, and how to identify which functions most meaningfully raise misuse risk in the wild.

Alongside research, enterprise-grade safety tools like Protect AI's Guardian platform~\cite{ref88} are becoming essential for securing open-weight model use in critical infrastructure and commercial applications. Guardian scans models for common threats and technical vulnerabilities, such as architectural backdoors, and runtime exploits. Its integration~\cite{ref89} with platforms like Hugging Face has already helped identify vulnerabilities in thousands of open-source models. While tools like this cannot prevent every instance of misuse, they provide a crucial layer of defence for organisations relying on open models---reducing exposure, flagging embedded risks, and supporting safer adoption.

For investors, supporting the growth and scaling of such technologies is not only an opportunity to meet rising demand in AI safety, but a way to accelerate the transition from experimental safeguards to standard, widely adopted infrastructure across the open-weight development pipeline. For instance, Protect AI is currently in the process of acquisition~\cite{ref90} with Palo Alto Networks, while another startup~\cite{ref91} focusing on securing AI systems also completed a successful investment round. These developments reflect strong investor confidence in AI safety-focused ventures and the expanding market for such solutions. That said, it is equally critical to ensure these technologies remain accessible to critical public infrastructure at affordable costs---not only to protect private sector deployments, but to bolster systemic resilience at a national and global level.

Finally, research collectives and transparency advocates must reckon with a central tension: full openness---including the release of weights, training data, and code---does not simply enhance reproducibility, it expands the attack surface. Initiatives like EleutherAI's~\cite{ref92} GPT models exemplify community-driven transparency and oversight, but typically lack formal release governance or systematic threat modelling. Releasing models alongside training data may aid scientific replication, but it can also enable more capable misuse by lowering the barrier to retraining or repurposing.

Funders and academic institutions should prioritise work that disentangles which components of AI systems contribute most to real-world risk, and under what conditions. This means investing not just in documentation, benchmarking, and interpretability tooling, but also in rigorous frameworks for release evaluation, red teaming pipelines, and sandboxed research access. By resourcing research into selective openness, safety-preserving transparency, and scalable misuse detection, the broader research ecosystem can ensure that openness serves not only innovation---but also security, accountability, and public trust.

With growing interest in open-weight models as public infrastructure, it is imperative that policymakers and industry leaders invest more in developing robust technical guardrails that can be coupled with advancing capabilities, ensuring long-term safety and security. These investments should focus not just on post-release mitigations but on building safety directly into the weights and distribution mechanisms of the models themselves.

\subsection{Enhance Societal Preparedness}
\label{sec:preparedness}

As open-weight advanced AI models become more capable, societal resilience must advance in parallel. These models, by their broad accessibility and adaptability, lower barriers to misuse and create new vulnerabilities across critical systems. Governance is essential to reducing risks at their source, but even the strongest safeguards cannot eliminate all threats. Preparing for the risks that will still arise is not a concession to failure---it is a recognition of the complexity of the challenge. Building defensive capabilities, strengthening detection systems, and raising public awareness are critical to managing the impacts of misuse when it occurs. The current stage of open-weight AI development offers a narrow window: societies can still prepare deliberately, while threats remain comparatively limited. This preparation must reinforce, not replace, sustained efforts to govern advanced AI responsibly.

While societal preparedness takes many forms~\cite{ref93} and will expand to new areas as advanced AI grows more capable, here are some key approaches we can implement today to build resilience for the future.

\subsubsection{Public and Private Institutions for AI Literacy}
\label{sec:literacy}

Public understanding of advanced AI capabilities must grow alongside their availability. As open-weight models become more widely deployed, both the general public and key institutions need a clearer grasp of what these systems can and cannot do. Public-facing literacy initiatives should help users distinguish between realistic threats and inflated narratives, while addressing practical concerns like AI-generated scams, fraud, and deepfakes. These efforts can draw from cybersecurity and media literacy precedents but must now be adapted for the unique attributes of general-purpose AI. Tools like AI Digest~\cite{ref94} can play a valuable role in surfacing new capabilities and informing civil society, not to raise alarm, but to help build informed expectations and support governance grounded in actual risk. This anticipatory mindset can be instrumental to identify and address newly emerging public safety aspects of advanced AI adoption, such as digital mental health~\cite{ref95}.

At the same time, specialised literacy programs must be developed for policymakers and professionals involved in critical infrastructure and security. These audiences require deeper understanding of how open-weight models could be exploited to escalate cyberattacks, or enable dual-use applications specifically in their sector and workstreams. A strong example of this approach is the EU AI Act's mandate~\cite{ref96} requiring deployers of AI systems to ensure sufficient AI literacy among their personnel. However, it is important that this mandate does not stay on the surface, but is comprehensive especially for critical infrastructure and services, including structured threat modelling, misuse escalation scenarios, and proactive measures to mitigate such risks.

\subsubsection{National Cybersecurity, CBRN and Critical Infrastructure Agencies}
\label{sec:national-security}

National security, cybersecurity, and biosecurity agencies must establish AI-specific incident-monitoring and response teams capable of detecting and countering model-enabled threats. Because open-weight AI is a universal technology, the plans these agencies develop should expand beyond their national AI landscape, and keep the global threat landscape in mind---linking domestic systems to cross-border intelligence exchanges, joint drills, and multilateral response protocols.

These teams should track misuse patterns such as AI-generated exploit code, model-assisted vulnerability scanning, dynamic cyber-attacks targeting critical infrastructure, and potential AI-enabled CBRN misuse. Structured national incident-reporting infrastructures must be built to capture, classify, and analyse these incidents systematically, providing early-warning capabilities rather than relying on retrospective analysis. Building resilient national reporting systems is also a necessary foundation for future international coordination. As emphasised by the OECD's common reporting framework for AI incidents~\cite{ref97}, interoperability between national systems will be critical, but domestic detection, classification, and response capabilities must come first.

Recent initiatives, such as the UK's 2025 AI Cybersecurity Code of Practice~\cite{ref98}, demonstrate how national investment in AI-specific security standards can support both domestic resilience and international norm-setting. Regular red-teaming exercises focused on open-weight model misuse must be conducted, simulating not only direct cyberattacks but also complex, multi-domain threat scenarios that blend AI-enabled tactics across information, infrastructure, and critical service domains.

National biosecurity and non-proliferation agencies should likewise strengthen protections around sensitive biological and chemical databases, tighten controls over dual-use research publications, and enhance oversight of laboratory equipment, precursor materials, and acquisition pathways. Scrutiny at these critical junctures is essential to prevent AI capabilities from translating into real-world physical threats. As an additional measure, governments could integrate biosecurity red-team exercises into existing national cybersecurity drills to ensure cross-domain readiness.

\subsubsection{International CBRN \& Biosecurity Coordination}
\label{sec:cbrn}

International oversight bodies such as the UN BioRisk Working Group~\cite{ref99} and the Organisation for the Prohibition of Chemical Weapons~\cite{ref100} can support information sharing, develop common evaluation benchmarks, and promote global coordination on AI-related CBRN risks. These organisations might have the power to convene AI researchers and technical advisers within their internal networks, such as relevant UN working groups, or they can coordinate across member countries to pool relevant resources---such as experts, rapid-response funding, necessary tooling or data sets. By harmonising threat-assessment methodologies and publishing open guidance on best-practice mitigations, these bodies help national agencies align their safeguards with emerging technical realities. Early capacity-building efforts like the OPCW-hosted AI and Chemical Safety and Security Management Workshop~\cite{ref101} illustrate the value of this approach and should evolve into a broader, more rigorous agenda with wide participation from member states.

\section{Conclusion}
\label{sec:conclusion}

Society deserves more than a binary choice between open and closed AI.

Open-weight models are a powerful driver of innovation, but their release also creates permanent and widely distributed access to advanced capabilities, some of which may be misused in ways that current safeguards are not equipped to prevent. In the current catch-up game between closed and open AI development, governance mechanisms remain underdeveloped, and the risks of misuse are growing less theoretical and more immediate.

The goal is not to shut the door on openness, but to ensure that we open it deliberately---with foresight, caution, and credible safeguards.

The steps outlined in this report reflect good starting points on how to get there. Technical safeguards must be designed to withstand real-world deployment---not just perform in idealised conditions. Release decisions must be tied to concrete, testable safety criteria---not aspirational principles alone. Developers, policymakers, and domain experts need coordinated processes and shared accountability---not disconnected efforts that create blind spots and loopholes. And the wider public must be prepared---not only shielded---through stronger institutional capacity, clearer communication, and a better understanding of how these models may reshape critical systems.

Taken together, these investments enable a tiered approach to open-weight release: one that supports safe access where it is possible, and defers release where it is not. The tiered system is not just a technical fix---it is a principled stance. If a model cannot be released safely, it must not be released at all. This is not a failure of openness, but a sign of responsible governance. Even a carefully designed approach may not guarantee perfect safety---but it offers an actionable path to significantly mitigate risk, grounded in today's knowledge, and designed for the challenges ahead, including those we may not get another chance to prepare for.

{\sloppy

}

\section*{How to Cite This Publication}

\begin{quote}
\"{O}zcan, B., Petropoulos, A., and Reddel, M. (2025). \textit{Beyond the Binary: Why Advanced AI Demands More Than the Open vs Closed Debate}. Centre for Future Generations. July 30, 2025.
\end{quote}

\noindent BibTeX:
\begin{verbatim}
@techreport{ozcan2025beyond,
  author    = {\"Ozcan, Beng\"usu and Petropoulos, Alex
               and Reddel, Max},
  title     = {Beyond the Binary: Why Advanced AI Demands
               More Than the Open vs Closed Debate},
  institution = {Centre for Future Generations},
  year      = {2025},
  month     = {July},
  url       = {}
}
\end{verbatim}

\noindent\textit{This publication was originally designed and optimised for web and published on cfg.eu. Minor formatting differences may appear in this version.}

\end{document}